\documentclass[article,nojss]{jss}
\usepackage{amsfonts,amstext,amsmath,bm,orcidlink,thumbpdf,lmodern}
\setkeys{Gin}{width=\textwidth}

\author{Achim Zeileis~\orcidlink{0000-0003-0918-3766}\\Universit\"at Innsbruck}
\Plainauthor{Achim Zeileis}

\title{Examining Exams Using Rasch Models and Assessment of Measurement Invariance}
\Shorttitle{Examining Exams}

\Abstract{
  Many statisticians regularly teach large lecture courses on statistics,
  probability, or mathematics for students from other fields such as business
  and economics, social sciences and psychology, etc. The corresponding exams
  often use a multiple-choice or single-choice format and are typically
  evaluated and graded automatically, either by scanning printed exams or via
  online learning management systems. Although further examinations of these
  exams would be of interest, these are frequently not carried out. For example
  a measurement scale for the difficulty of the questions (or items) and the
  ability of the students (or subjects) could be established using psychometric
  item response theory (IRT) models. Moreover, based on such a model it could be
  assessed whether the exam is really fair for all participants or whether
  certain items are easier (or more difficult) for certain subgroups of students.

  Here, several recent methods for assessing \emph{measurement invariance} and
  for detecting \emph{differential item functioning} in the Rasch IRT
  model are discussed and applied to results from a first-year mathematics exam
  with single-choice items. Several categorical, ordered, and numeric covariates
  like gender, prior experience, and prior mathematics knowledge are available
  to form potential subgroups with differential item functioning. Specifically,
  all analyses are demonstrated with a hands-on \proglang{R} tutorial using the
  \emph{psycho*} family of \proglang{R}~packages (\pkg{psychotools},
  \pkg{psychotree}, \pkg{psychomix}) which provide a unified approach to
  estimating, visualizing, testing, mixing, and partitioning a range of
  psychometric models.
  
  The paper is dedicated to the memory of Fritz Leisch (1968--2024) and his
  contributions to various aspects of this work are highlighted.
}

\Plainkeywords{multiple choice, item response theory, differential item functioning, psychometrics, R}

\Address{
  Achim Zeileis\\
  Universit\"at Innsbruck\\
  Department of Statistics\\
  Universit\"atsstr.~15\\
  6020 Innsbruck, Austria\\
  E-mail: \email{Achim.Zeileis@R-project.org},\\
  URL: \url{https://www.zeileis.org/}
}

\begin{document}

\section{Introduction} \label{sec:intro}

\subsection{Large-scale exams}

Statisticians often teach large lecture courses with introductions to
statistics, probability, or mathematics in support of other curricula such as
business and economics, social sciences, psychology, etc. Due to the large
number of students and possibly also of lecturers who teach lectures and/or
tutorials in parallel, it is often necessary to rely on exams and other
assessments based on large pools of so-called closed (as opposed to open-ended)
items, i.e., exercises which can be evaluated and graded automatically.

\pagebreak

The most widely-used item type for such assessments are multiple-choice (also
known as multiple-answer) or single-choice exercises. But due to the widespread
adoption of learning management systems such as Moodle \citep{Moodle},
Canvas \citep{Canvas}, or Blackboard \citep{Blackboard}, especially since the
Covid-19 pandemic, other item types are also being used increasingly.
Often the evaluation is binary (correct vs.\ incorrect)
but scores with partial credits for partially correct items are also
frequently used.

\subsection{Examining exams}

Traditionally, mostly simple summary statistics have been used for the results
from such large-scale exams, e.g., the proportion of students who correctly solved
the different items and the number of items solved per student.
However, recently there has also been increasing interest in so-called
\emph{learning analytics} which connects the results from different exams
or assessments as well as covariates such as the field and duration of the
study, prior knowledge from previous courses, etc., in order to better understand
and shape the learning environments for the students
\citep[see][for an overview and further references]{Wiki+LearningAnalytics}.

However, in Austria, to the best of our knowledge, it is still not common to
apply standardized and/or automated psychometric assessments to exam results. A
notable exception is the multiple-choice monitor at WU Wirtschaftsuniversit\"at
Wien introduced by \cite{Nettekoven+Ledermueller:2012,Nettekoven+Ledermueller:2014}.
In addition to various exploratory techniques they also employ probabilistic
statistical models from psychometrics to gain more insights into exam results.
More specifically, they use models from item response theory
\citep[IRT,][]{Fischer+Molenaar:1995,VanDerLinden+Hambleton:1997}, including the
Rasch model \citep{Rasch:1960} which is also employed in the analysis of international
educational attainment studies such as PISA (Programme for International Student
Assessment, \url{https://www.oecd.org/pisa/}).

\subsection{Measurement invariance in IRT}

Based on an exam's item responses, IRT models can estimate various quantities of
interest, most importantly the \emph{ability} of the individual students and
the \emph{difficulty} of the different items (or exercises). A fundamental assumption
is that the models' parameters are invariant across all observations,
which is also known as \emph{measurement invariance}
\citep[see][for an early overview in psychometrics]{Horn+McArdle:1992}.
Otherwise observed differences in the items solved cannot be reliably
attributed to the latent variable that the model purports to measure.

Typical sources for violation of measurement invariance in IRT models are
\emph{multidimensionality} (i.e., more than one latent variable instead of a single
ability) or \emph{differential item functioning} \citep[see][]{Debelak+Strobl:2024}. The
latter refers to the situation where the same item can be relatively easier or more
difficult (compared to the remaining items) for different students, despite having the
same latent ability.

\subsection{Our contribution}

In Section~\ref{sec:data} we introduce a data set from one of our own mathematics
courses, containing binary responses (correct vs.\ incorrect) from 13~items
in an end-term exam of an introductory mathematics course for economics and
business students. In Section~\ref{sec:irt} the Rasch IRT model is briefly
introduced, fitted to the data, and interpreted regarding the items' difficulties
and students' abilities. Subsequently, in Section~\ref{sec:dif} various methods
for capturing violations of measurement invariance are applied: (1)~Classical
two-sample comparisons of two exogenously given groups along with modern
methods for anchoring the item difficulty estimates. (2)~Rasch trees based on
generalized measurement invariance tests for data-driven detection of subgroups
affected by DIF. (3)~Rasch finite mixture models as an alternative way of
data-driven characterization of DIF clusters. Section~\ref{sec:discussion}
wraps up the paper with a discussion and the epilogue Section~\ref{sec:epilogue}
concludes the paper by highlighting Fritz Leisch's influence on different aspects
of this work.

In all sections, emphasis is given to the hands-on application of the methods
in \proglang{R} \citep{R} -- notably using the packages \pkg{psychotools}
\citep{psychotools},
\pkg{psychotree} \citep{psychotree}, and \pkg{psychomix} \citep{psychomix}
-- along with the practical insights about the analyzed exam.

\section{Data: Mathematics 101 at Universit\"at Innsbruck} \label{sec:data}

The data considered for examination in the following sections come from the
end-term exam in our ``Mathematics 101'' course for business and economics
students at Universit\"at Innsbruck. This is a course in the first semester
of the bachelor program and it is attended by about 600--1,000 (winter) 
or 200--300 (summer) students per semester.

Due to the large number of students in the course, there are frequent online
tests carried out in the university's learning management system OpenOlat
\citep{OpenOlat} as part of the tutorial groups, along with two written exams.
All assessments are conducted with support from \proglang{R} package \pkg{exams}
\citep{Gruen+Zeileis:2009,Zeileis+Umlauf+Leisch:2014} which allows to
automatically generate a large variety of similar exercises and render these
into many different output formats.

In the following the individual results from an end-term exam are analyzed
for 729 students (out of 941 that had registered at the beginning of the
semester). The exam consisted of 13~single-choice items with five answer alternatives,
covering the basics of analysis, linear algebra, and financial mathematics.
Due to the high number of participants, the exam was conducted with two groups,
back to back, using partially different item pools (on the same topics). All
students had individual versions of their items generated via \proglang{R}/\pkg{exams}.
Correctly solved items yielded 100\% of points associated with an exercise.
Items without correct solution can either be unanswered (0\%) or have an
incorrect answer ($-25\%$). In the following, the item responses are treated
as binary (correct vs.\ not correct).

The data are available in the \proglang{R} package \pkg{psychotools} as
\code{MathExam14W} where \code{solved} is the main variable of interest.
This is an object of class \code{itemresp} which is internally essentially
a $729 \times 13$ matrix with binary 0/1 coding plus some metainformation.
In addition to the item responses, there are a number of covariates of interest:
\begin{itemize}
  \item \code{group}: Factor for group (\code{1} vs.\ \code{2}).
  \item \code{tests}: Number of previous online exercises solved (out of 26).
  \item \code{nsolved}: Number of exam items solved (out of 13).
  \item \code{gender}, \code{study}, \code{attempt}, \code{semester}, \dots
\end{itemize}
For a first overview, we load the package and data. Then we exclude those
participants with the extreme scores of 0 and 13, respectively, because
these students do not discriminate between the items (either none solved or
all solved). The \proglang{R} code below employs the \code{print()} and 
\code{plot()} methods for \code{itemresp} objects by printing the first couple of item
responses and visualizing the proportion of correct responses per item.
\begin{Schunk}
\begin{Sinput}
R> library("psychotools")
R> data("MathExam14W", package = "psychotools")
R> mex <- subset(MathExam14W, nsolved > 0 & nsolved < 13)
R> head(mex$solved)
\end{Sinput}
\begin{Soutput}
[1] {1,1,1,0,1,1,0,1,1,0,1,1,0} {1,1,1,1,0,0,0,1,1,1,1,1,1}
[3] {0,0,1,0,0,1,0,1,1,0,0,1,0} {0,1,0,1,1,1,1,1,1,1,1,1,1}
[5] {1,0,0,1,1,0,0,0,0,1,1,0,1} {1,0,0,1,1,0,0,0,1,1,0,0,0}
\end{Soutput}
\begin{Sinput}
R> plot(mex$solved)
\end{Sinput}
\end{Schunk}
\begin{figure}[t!]
\includegraphics{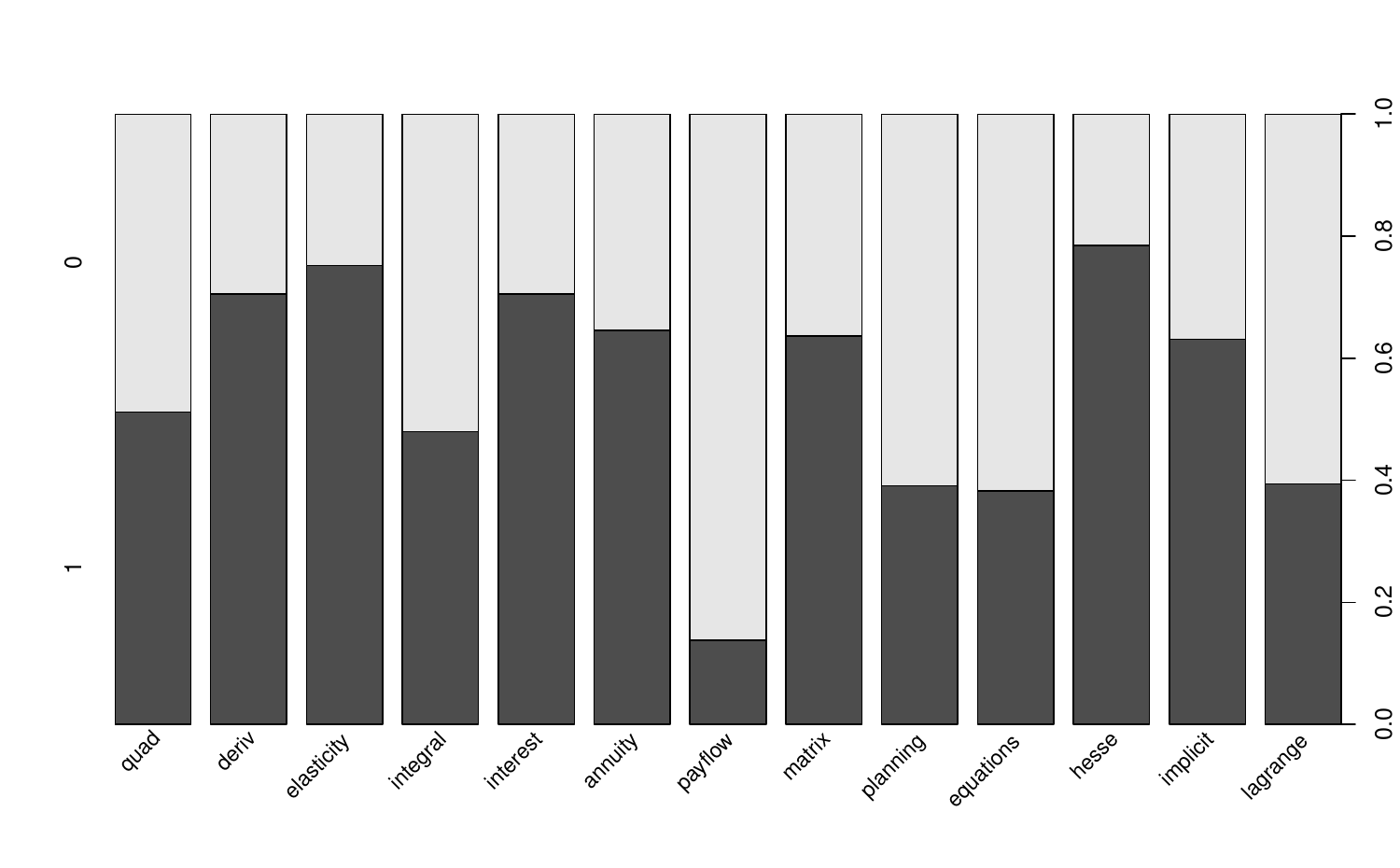}
\caption{Bar charts with relative frequencies of items solved correctly
(1, dark gray) or not solved correctly (0, light gray).
The item labels on the $x$-axis briefly indicate the topic of each item.}
\label{fig:itemresp}  
\end{figure}
The resulting plot is shown in Figure~\ref{fig:itemresp}, highlighting
that most items have been solved correctly by about 40 to 80 percent of the
participants. The main exception is the payflow exercise (for which a
certain integral had to be computed) which was solved correctly by less
than 15 percent of the students.

\section{IRT with the Rasch model} \label{sec:irt}

While the exploratory analysis of the item responses is already interesting,
item response theory \citep[IRT,][]{Fischer+Molenaar:1995,VanDerLinden+Hambleton:1997}
provides us with more refined methods to infer the underlying latent trait
captured by the exam (here: ability in basic mathematics). In particular,
it allows to align the items' difficulties and the students' abilities on
the same scale for the latent trait.

Here, we employ the basic Rasch model \citep{Rasch:1960} for modeling
binary items $y_{ij} \in {0, 1}$ based on $i = 1, \dots, n$ persons and
$j = 1, \dots, m$ items. The Rasch model then aligns a person's
ability $\theta_i$ and an item's difficulty $\beta_j$ using a logistic model.
\begin{eqnarray*}
  \pi_{ij} & = & \text{Pr}(y_{ij} = 1)\\
  \text{logit}(\pi_{ij}) & = & \theta_i - \beta_j
\end{eqnarray*}
This means that the probability $\pi_{ij}$ that person~$i$ solves item~$j$
correctly is linked to the difference of ability and difficulty through
a logit link where $\text{logit}(\pi) = \log\{\pi/(1 - \pi)\}$.

Most statistical audiences will be familiar with fitting and interpreting
logistic regression models for binary responses, but there are two aspects in
the Rasch model that require special attention: First, full maximum
likelihood estimation of all model parameters (also known as joint maximum
likelihood) is inconsistent because the number of parameters also diverges
as the number of observations increases to infinity (either by increasing
the number of persons $n$ or the number of items $m$). Second, the model
parameters $\theta_i$ and $\beta_j$ are only defined up to an additive
constant because only their difference is identified and adding the same
constant to both parameter groups would cancel out.

The first issue is usually solved by one of two approaches: Either it is
assumed that the ability parameters come from a normal distribution and then
the item parameters can be estimated by marginal maximum likelihood. Or,
alternatively, the abilities for person~$i$ are conditioned on the so-called sum scores
(the number of items solved by that person) and conditional maximum likelihood is used
\citep{Fischer+Molenaar:1995}. Throughout this paper we employ conditional
maximum likelihood estimation.

The second issue is solved by fixing a zero point on the latent trait scale,
e.g., by fixing the sum of all item parameters to zero. As long as only a single
model is fitted to all persons, the specific reference point is usually not of much
practical relevance. However, it is crucial when comparing the item parameter
estimates from two or more subgroups of persons. This will be revisited in
the next section. For now, we restrict the sum of item parameters to zero.

\begin{figure}[t!]
\includegraphics[trim=0 5 0 20, clip]{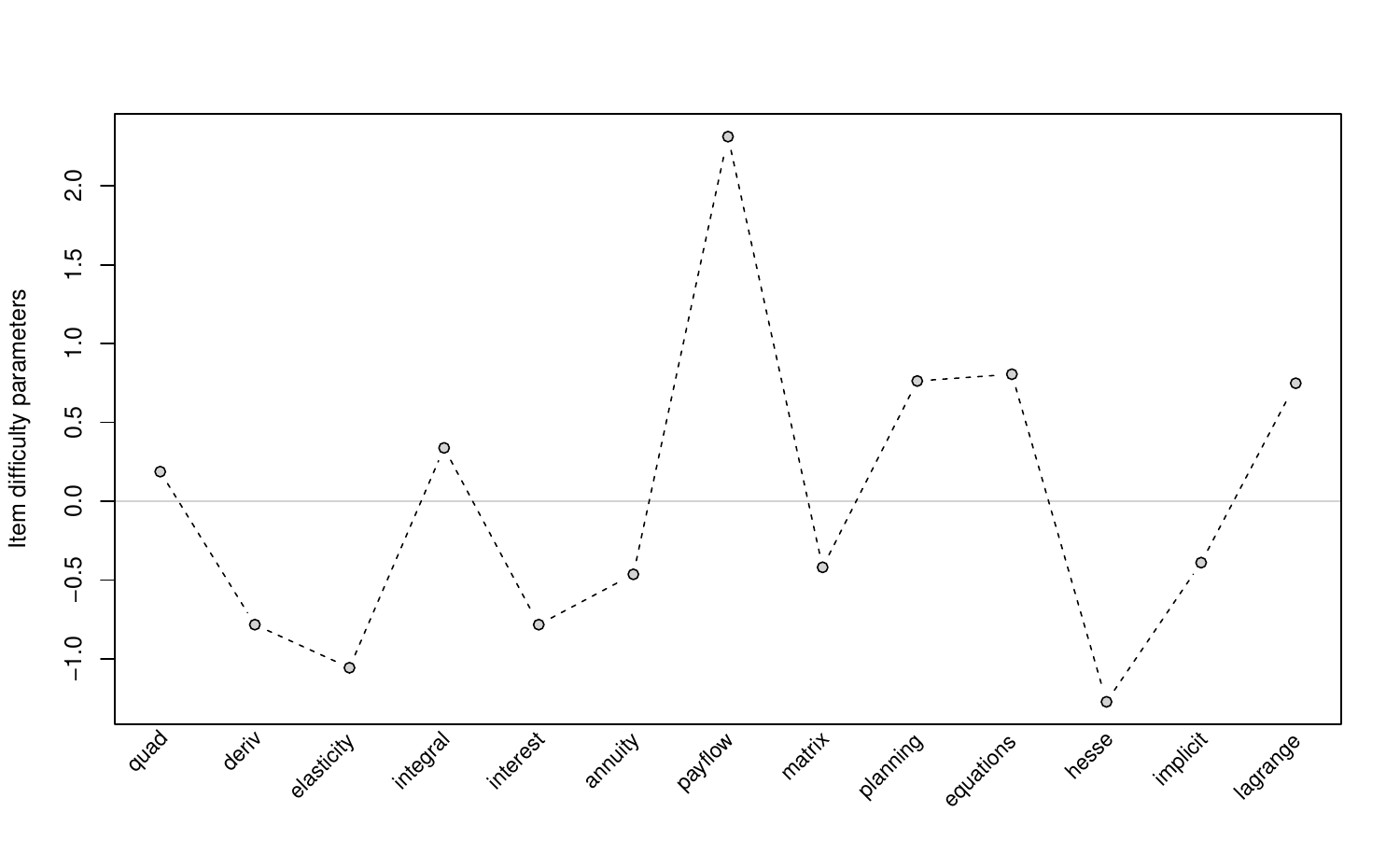}

\includegraphics{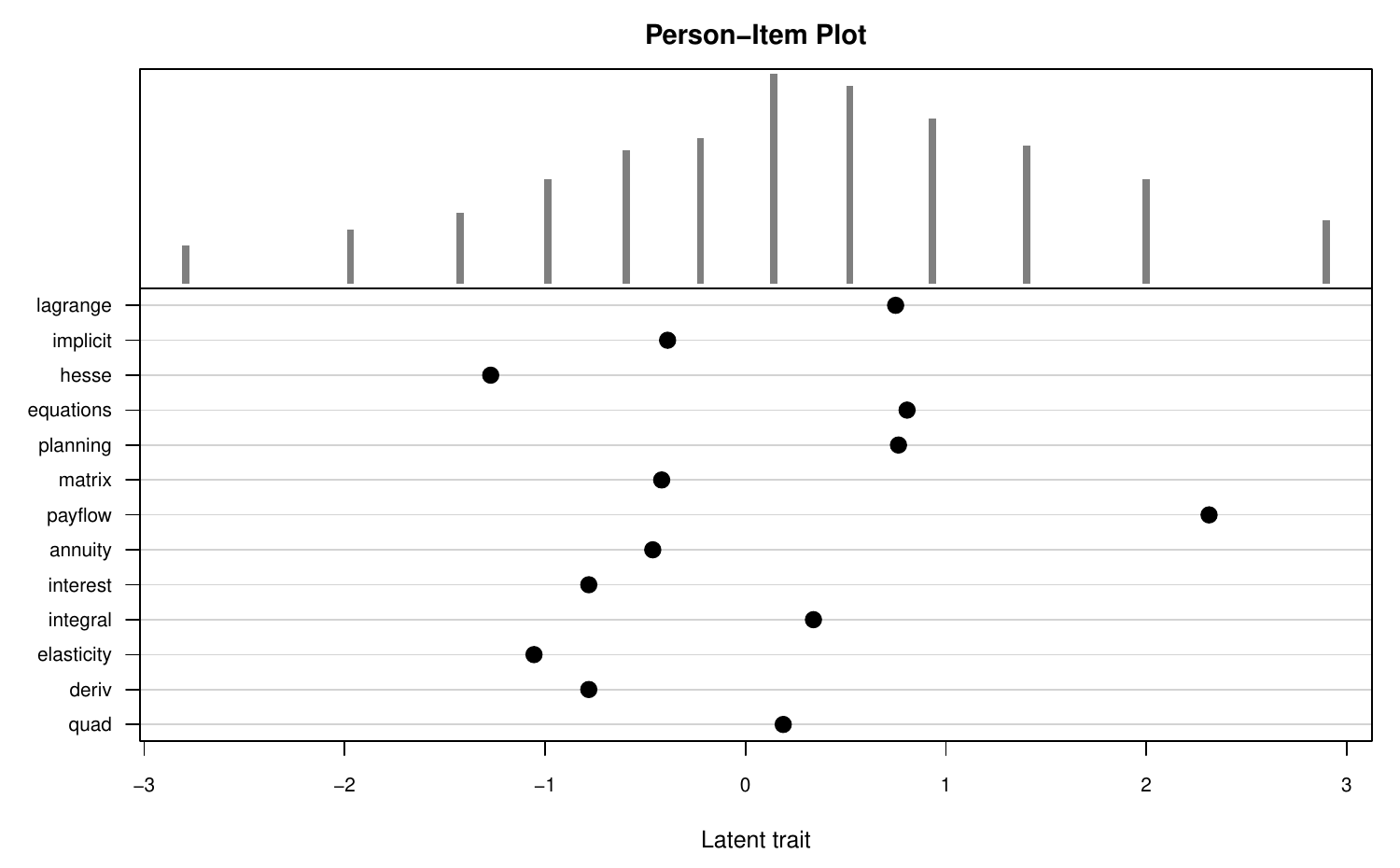}
\caption{Top: Profile plot of estimated item parameters (difficulties $\hat \beta_j$).
Bottom: Person-item plot of estimated person and item parameters (abilities) $\hat \theta_i$ and
$\hat \beta_i$.}
\label{fig:raschmodel}
\end{figure}

In \proglang{R} a number of packages are available to estimate various
IRT models, including the Rasch model
\citep[see][for an introduction]{Debelak+Strobl+Zeigenfuse:2022}. The most
popular ones are probably \pkg{eRm} \citep{Mair+Hatzinger:2007} and
\pkg{mirt} \citep{Chalmers:2012}. Here, we employ the implementation in
package \pkg{psychotools} \citep{psychotools} instead because it provides
a particularly rich toolbox of methods for assessing measurement invariance
which we leverage in the next section.

Fitting a Rasch model with \pkg{psychotools} can be carried out with
the function \code{raschmodel()} based on a binary item response matrix. For the resulting
model class a number of methods are available, e.g., extracting the item
and person parameters via \code{itempar()} and \code{personpar()}, respectively.
These parameters can then be displayed using the \code{plot()} methods
in various ways, e.g., just the ``profile'' of item parameters
(Figure~\ref{fig:raschmodel}, top) or a so-called person-item plot
(Figure~\ref{fig:raschmodel}, bottom) which displays boths sets of parameters
on the same scale.
\begin{Schunk}
\begin{Sinput}
R> mr <- raschmodel(mex$solved)
R> plot(mr, type = "profile")
R> plot(mr, type = "piplot")
\end{Sinput}
\end{Schunk}
Qualitatively, both panels in Figure~\ref{fig:raschmodel} show a similar pattern
as the empirical proportions from Figure~\ref{fig:itemresp}. However, due to the
latent logistic scale the most difficult item (payflow) and the easiest item
(hesse) are brought out even more clearly. Also the majority of the item difficulties
are close to the median ability in this sample. Thus, the exam discrimenates more
sharply at the median difficulty and less sharply in the tails at very high
or very low ability.

\section{Assessment of measurement invariance} \label{sec:dif}

The interpretation of the Rasch model parameters from the previous section is
only reliable if all item difficulties $\beta_j$ are indeed the same for all students
in the sample. If this is not the case, differences in the item responses would not
necessarily be caused by differences in mathematics ability. The fundamental assumption that
the item parameters are constant across all persons is a special case of
so-called measurement invariance. And a violation of this assumption is known as
differential item functioning \citep[DIF,][]{Debelak+Strobl:2024}, i.e., some item(s)
is/are relatively easier for some subgroup of persons compared to others.

To check whether measurement invariance is violated and whether there is differential
item functioning, it needs to be checked whether the item parameters differ across
persons \citep{Merkle+Zeileis:2013}, e.g., along with some available covariates.
For example, item difficulty might vary along with age or some items might be more
difficult for non-native speakers etc. Note that this is \emph{not} the situation
where a persons' ability varies along with some covariates because this would imply that
for such a person \emph{all} items are relatively more or less difficult. In contrast,
DIF refers to the situation where this occurs only for some item(s).

In the following, we consider three different methods for assessing DIF: First, if the
potential subgroups affected by DIF are known, it is possible to test for item parameter
differences between these groups using the classical likelihood ratio, score, or Wald tests
with suitable item anchoring \citep{Glas+Verhelst:1995}. Second, generalizations of the
score test can also assess changes along continuous or ordinal variables
\citep{Merkle+Zeileis:2013,Merkle+Fan+Zeileis:2014} and trees can repeat these
tests recursively for different variables to form subgroups in a data-driven way
\citep{Strobl+Kopf+Zeileis:2015}. Third, finite mixtures of Rasch models are a 
general strategy to test for violations of measurement invariance, even when there are
no covariates \citep[originally proposed by][]{Rost:1990}. But it is also possible
to inform the selection of subgroups in Rasch mixture models by covariates
\citep[see][]{Frick+Strobl+Leisch:2012}.

\subsection{Classical tests with reference and focal groups}

In the examination of our mathematics exam results, the obvious first question in a
DIF setting is whether the item parameters for the first and the second group differ.
Recall that the two groups took the exam in back to back sessions and the item categories
were the same for all 13~items. However, different concrete items were used from these
categories for the first and the second group. Thus, it is natural to suspect that at
least some of the items might have different difficulties.

The strategy for this is straightforward: We split the data into reference and focal groups
(here groups~1 and~2) and then assess the stability of selected parameters across the groups
by means of standard tests. Especially, the likelihood ratio (LR) test is easy to compute.
Its test statistic is twice the difference between the full-sample likelihood and the
overall likelihood from the two subgroups. Using the \pkg{psychotools} functionality
the LR statistics can be obtained ``by hand'' via:
\begin{Schunk}
\begin{Sinput}
R> mr1 <- raschmodel(subset(mex, group == 1)$solved)
R> mr2 <- raschmodel(subset(mex, group == 2)$solved)
R> -2 * as.numeric(logLik(mr) - (logLik(mr1) + logLik(mr2)))
\end{Sinput}
\begin{Soutput}
[1] 264.9577
\end{Soutput}
\end{Schunk}
This shows that the conditional likelihood of the model can be improved a lot by splitting
into groups~1 and~2 and that the LR test statistic is much larger than the 95\% critical value
$21.0$ from
the $\chi_{12}^2$ null distribution (because the split necessitates the estimation of 12~additional item parameters).
Very similar results would be obtained when using the Wald statistic $249.4$ or the
score (or Lagrange multiplier) statistic $260.8$.

\begin{figure}[t!]
\includegraphics[trim=0 5 0 50, clip]{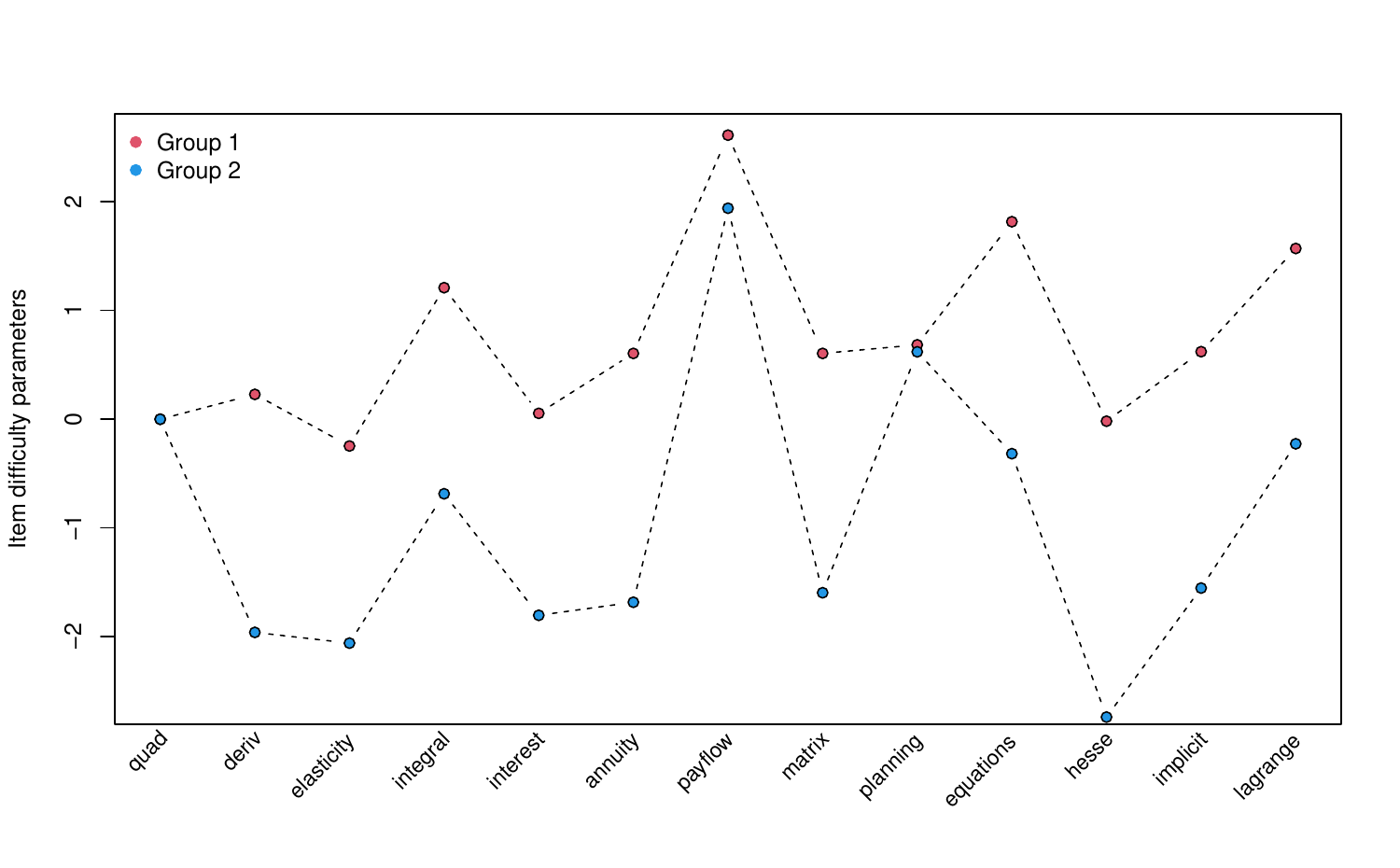}

\includegraphics[trim=0 5 0 50, clip]{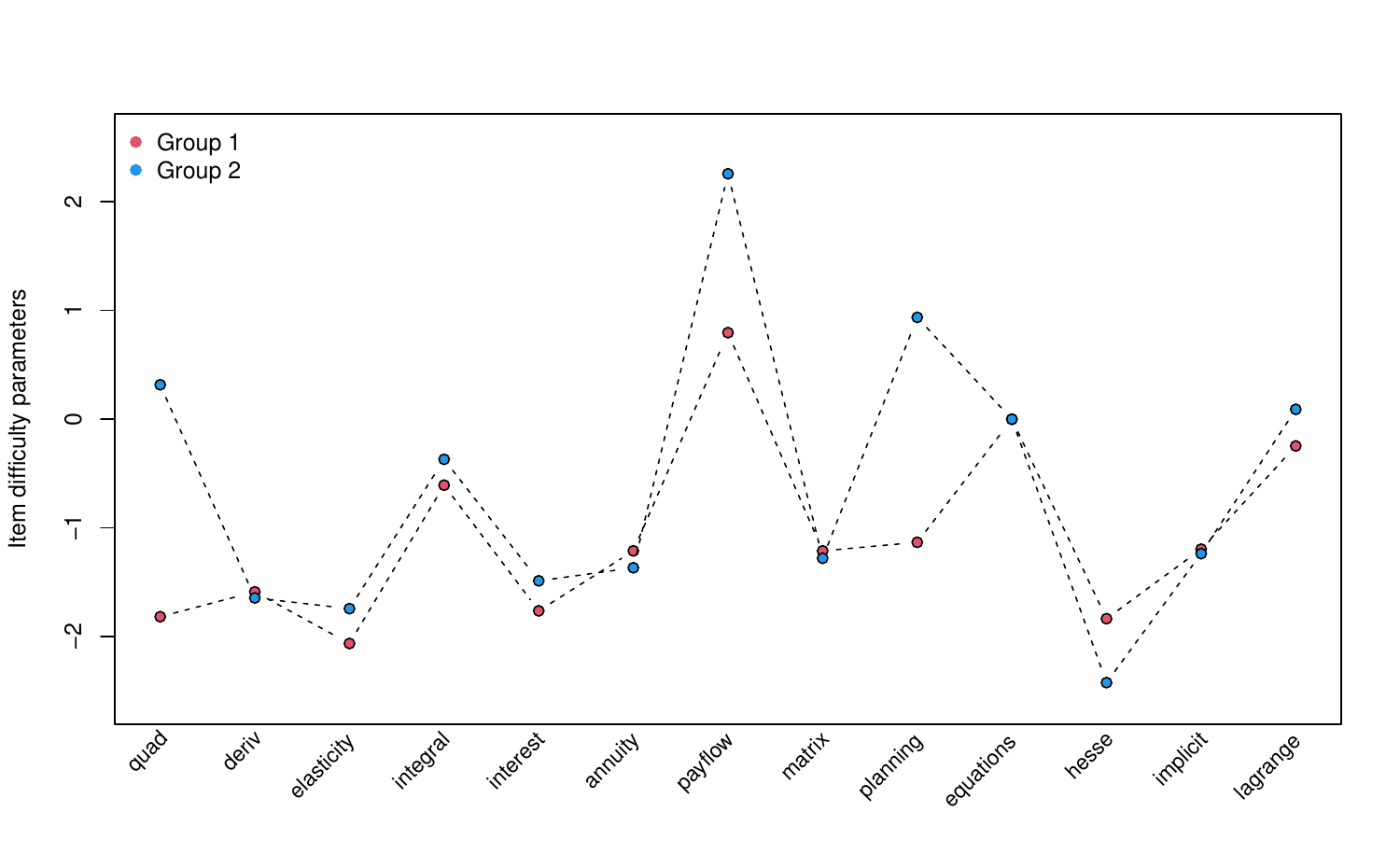}
\caption{Top: Profile plot of estimated item parameters (difficulties $\hat \beta_j$) with the constraint $\hat \beta_1 = 0$.
Bottom: Profile plot with the constraint $\hat \beta_{10} = 0$.}
\label{fig:profile-dif}
\end{figure}

Given that there is such strong evidence for DIF between the two groups, the
natural next question is: Which items ``cause'' this DIF? A natural strategy for answering
this question is looking at the item-wise Wald tests. This is simply the difference
between the item parameter estimate for item $j$ from the reference and the focal
group, scaled by the corresponding standard error:
\[ \displaystyle
t_j = \frac{\hat{\beta}_{j}^{\mbox{\tiny{ref}}} - \hat{\beta}_{j}^{\mbox{\tiny{foc}}}}
{\sqrt{ \widehat{\mbox{Var}}( \hat{\beta}^{\mbox{\tiny{ref}}})_{j,j} + 
\widehat{\mbox{Var}}( \hat{\beta}^{\mbox{\tiny{foc}}})_{j,j}
}}.
\]
However, there is one important caveat: As the reference point of zero from the
two scales is arbitrary, we need to find a way to ``anchor'' the scales from 
the two groups. Typically, this is done by selecting either a single item
or groups of items whose (mean of) parameters are used as the zero reference point
\citep{Woods:2009}. Unfortunately, though, the selection of the anchor item(s)
has a large influence on the results of the item-wise Wald tests and there is
no obvious election strategy. This is illustrated in Figure~\ref{fig:profile-dif}
which shows anchoring with item~1 (quad, top panel) and item~10 (equations, bottom panel).
The top panel is essentially generated as follows:
\begin{Schunk}
\begin{Sinput}
R> plot(mr1, parg = list(ref = 1), col = 2, ylim = c(-2.6, 2.6))
R> plot(mr2, parg = list(ref = 1), col = 4, add = TRUE)
\end{Sinput}
\end{Schunk}
The bottom panel analogously uses \code{ref = 10}.

With the anchoring in the top panel one would have to conclude that
almost all items differ in their difficulties except the fixed item~1 (quad)
and item~9 (planning). However, a completely different picture emerges
for the anchoring with item~10 in the bottom panel. Note that the patterns
within each profile are exactly the same, they are just both shifted vertically
so that the item parameter for item~10 is zero in both groups. With this
anchoring, only items~1 (quad), 7 (payflow), and~9 (planning) differ more
substantially whereas the remaining items function very similarly.

The latter interpretation is much more plausible and can also be backed up
by looking at the content of item~10. This was the same item template across
both groups, namely solving a system of three linear equations, where all students
just had somewhat different coefficients within the equations. Thus, this
item is a natural candidate for being an anchor item as it can be expected
to function the same for both groups.

To further validate this conclusion, we also consider a data-driven choice
of the anchor item(s). A lot of different strategies have been proposed
for this purpose \citep[for an overview see][]{Kopf+Zeileis+Strobl:2015,Kopf+Zeileis+Strobl:2015a}
which are based on the basic idea to select some DIF-free anchor
item(s) to be able to identify items truly associated with DIF. Given that
this is a kind of ``chicken or the egg'' dilemma, it is not so easy to solve
and there is no consensus as to what is the best way to do so. Here, we
use the recently proposed strategy by \cite{Strobl+Kopf+Kohler:2021},
which has been shown to have a simple motivation and good empirical performance.
This strategy selects the anchor item that minimizes
the inequality (as measured by the so-called Gini coefficient) among the
item parameter differences. Here, this yields item~12 (implicit) as the
best anchor item. However, the results are very similar to those from
using item~10, leading to qualitatively the same insights.

\begin{figure}[t!]
\includegraphics{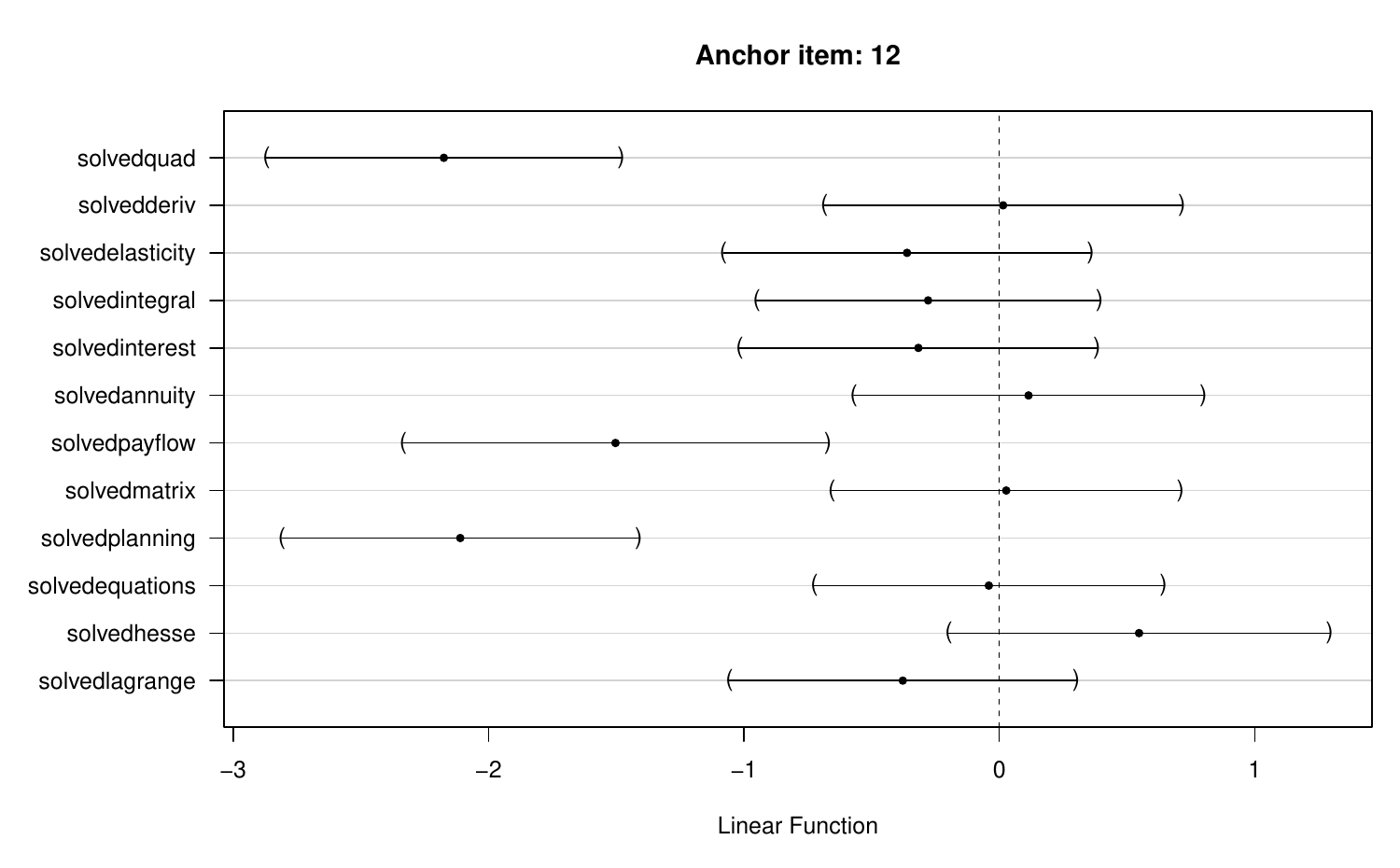}
\caption{Point estimates and corresponding simultaneous confidence intervals for
all item parameter differences between group~1 minus group~2, except for anchor item~12
(implicit) whose item parameter estimates are restricted to zero in both groups.}
\label{fig:anchortest}
\end{figure}

In \pkg{psychotools} this anchor strategy is the default in the \code{anchortest()}
function. We call it and indicate that we want to test for DIF in the
\code{solved} items by \code{group} and adjust for multiple testing
(across the 12 estimated item parameters) with the single-step method
from \pkg{multcomp} \citep{Hothorn+Bretz+Westfall:2008}.
\begin{Schunk}
\begin{Sinput}
R> ma <- anchortest(solved ~ group, data = mex, adjust = "single-step")
R> plot(ma$final_tests)
\end{Sinput}
\end{Schunk}
The visualization from the \code{plot()} method (see Figure~\ref{fig:anchortest})
does not show the two item profiles (as in Figure~\ref{fig:profile-dif}) but
shows confidence intervals for all item parameter differences (except the anchor
item whose difference is fixed to zero). This brings out clearly that items~1
(quad), 7 (payflow), and~9 (planning) are significantly more difficult for group~2
than for group~1, relative to the remaining items which do not significantly
violate the measurement invariance assumption.

In short, due to the DIF found, the results for groups~1 and~2 cannot be compared
completely fairly. Group~2 was somewhat disadvantaged by getting the more difficult
questions. Moreover, the question remains if there is even further DIF within these
two groups which might be associated with some of the other available covariates.

\subsection{Rasch trees}

The comparison between a reference and a focal group using classical tests is
very nice and straightforward for those familiar with maximum likelihood inference.
However, the main drawback is that the potential
subgroups have to be formed in advance. This is particularly inconvenient
when testing for DIF along continuous variables which, in practice, are often
categorized into groups in an ad hoc way (e.g., splitting at the median).
Also, when considering DIF in ordinal variables, the ordering of categories
is often not exploited, thus assessing only if at least one group differs from
the others.

To address these problems, \cite{Merkle+Zeileis:2013} and \cite{Merkle+Fan+Zeileis:2014}
have proposed generalizations of the classical score test which allow to assess
parameter instabilities along numeric and ordinal covariates, respectively.
These tests require no split into subgroups but just rely on the ordering
implied by the covariate.

\begin{figure}[t!]
\includegraphics[trim=0 5 0 10, clip]{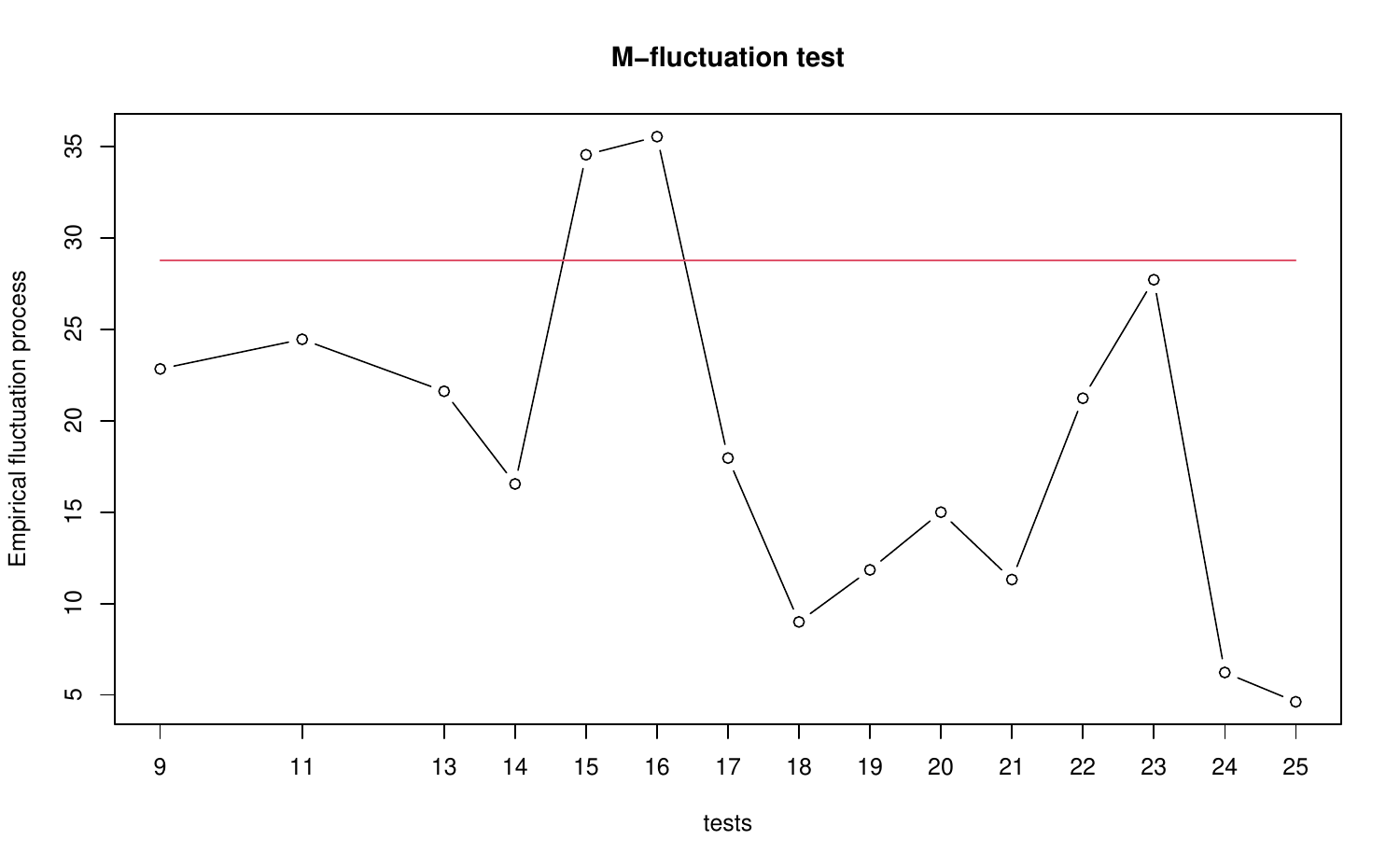}
\caption{Sequence of score (or Lagrange multiplier) test statistics for splits in
the number of points in the online tests (within the first group). The horizontal
red line is the 95 percent critical value for the maximum of the score statistics.}
\label{fig:maxlmo}
\end{figure}

Here, we first illustrate one of their tests for a numeric covariate by testing
for DIF along \code{tests} in \code{group}~1. Recall that \code{tests} is the number
of points obtained in the online tests throughout the semester and thus captures a
certain skill level and amount of preparation prior to the exam. The test statistic used
is simply the maximum of the score (or Lagrange multiplier) statistics for each
possible split in \code{tests}. Figure~\ref{fig:maxlmo} shows the sequence of score
statistics beginning with the one where students with $\mathtt{tests} \le 9$ form the
reference group and $\mathtt{tests} > 9$ the focal group. Then the same kind of
statistics are computed for splits at $11$~points in the \code{tests}, for $13$, etc.
Eventually, the null hypothesis has to be rejected if the maximum of theses tests becomes
larger than the corresponding critical value, which, of course, has to be adjusted
for the fact that the maximum of multiple test statistics is considered here.
Figure~\ref{fig:maxlmo} shows the 95 percent critical value as the horizontal red
line which is clearly exceeded by the sequence of test statistics. The highest
test statistics, and thus the highest amount of DIF, is obtained for a split
at $\mathtt{tests} = 16$.

In \proglang{R}, this maximum score test can be
carried out using the \pkg{strucchange} package \citep{Zeileis+Leisch+Hornik:2002}
which produces the Figure~\ref{fig:maxlmo} and also obtains a $p$-value by 
simulation.
\begin{Schunk}
\begin{Sinput}
R> library("strucchange")
R> mex1 <- subset(mex, group == 1)
R> sctest(mr1, order.by = mex1$tests, vcov = "info", functional = "maxLMo",
+    plot = TRUE)
\end{Sinput}
\begin{Soutput}
	M-fluctuation test

data:  mr1
f(efp) = 35.543, p-value = 0.005373
\end{Soutput}
\end{Schunk}
The result indicates that within the first group there is further evidence for DIF.
Students who performed rather poorly in the previous online tests have a different item
profile. More details will be provided below.

Now the natural next step is again to form subgroups for $\mathtt{tests} \le 16$
and $\mathtt{tests} > 16$, respectively, inspect the corresponding item parameter
profiles, and possibly test for further DIF. This strategy can be formalized using
so-called model-based recursive partitioning \citep{Zeileis+Hothorn+Hornik:2008}
in combination with Rasch models. The resulting Rasch trees
\citep{Strobl+Kopf+Zeileis:2015} are constructed recursively in the following steps:
\begin{itemize}
\setlength{\itemsep}{2pt}\setlength{\parskip}{2pt}
  \item Fit a Rasch model to the current subsample.
  \item Asess DIF along all covariates of interest (applying a Bonferroni adjustment for multiple testing).
  \item Split with respect to the covariate with the smallest significant $p$-value.
  \item Select split point by maximizing the log-likelihood.
  \item Continue until there are no more significant instabilities (or the sample is too small).
\end{itemize}
This strategy can be applied relatively easily using the \pkg{psychotree} package
which combines the model-based recursive partitioning infrastructure from \pkg{partykit}
with the IRT models from \pkg{psychotools}. To apply the method to the mathematics
exam data, we treat all numeric variables as ordinal because they have relatively
few distinct levels.
\begin{Schunk}
\begin{Sinput}
R> library("psychotree")
R> mex <- transform(mex,
+    tests    = ordered(tests),
+    nsolved  = ordered(nsolved),
+    attempt  = ordered(attempt),
+    semester = ordered(semester)
+  )
R> mrt <- raschtree(solved ~ group + tests + nsolved + gender +
+    attempt + study + semester, data = mex,
+    vcov = "info", minsize = 50, ordinal = "L2", nrep = 1e5)
\end{Sinput}
\end{Schunk}
The arguments in the last line specify that the information matrix is used as the
estimate for the variance-covariance matrix, that each subgroup must have at least
50~persons, and that the maximum score test is used for the ordinal covariates
with \code{nrep} replications in the simulated $p$-values.

\begin{figure}[t!]
\includegraphics{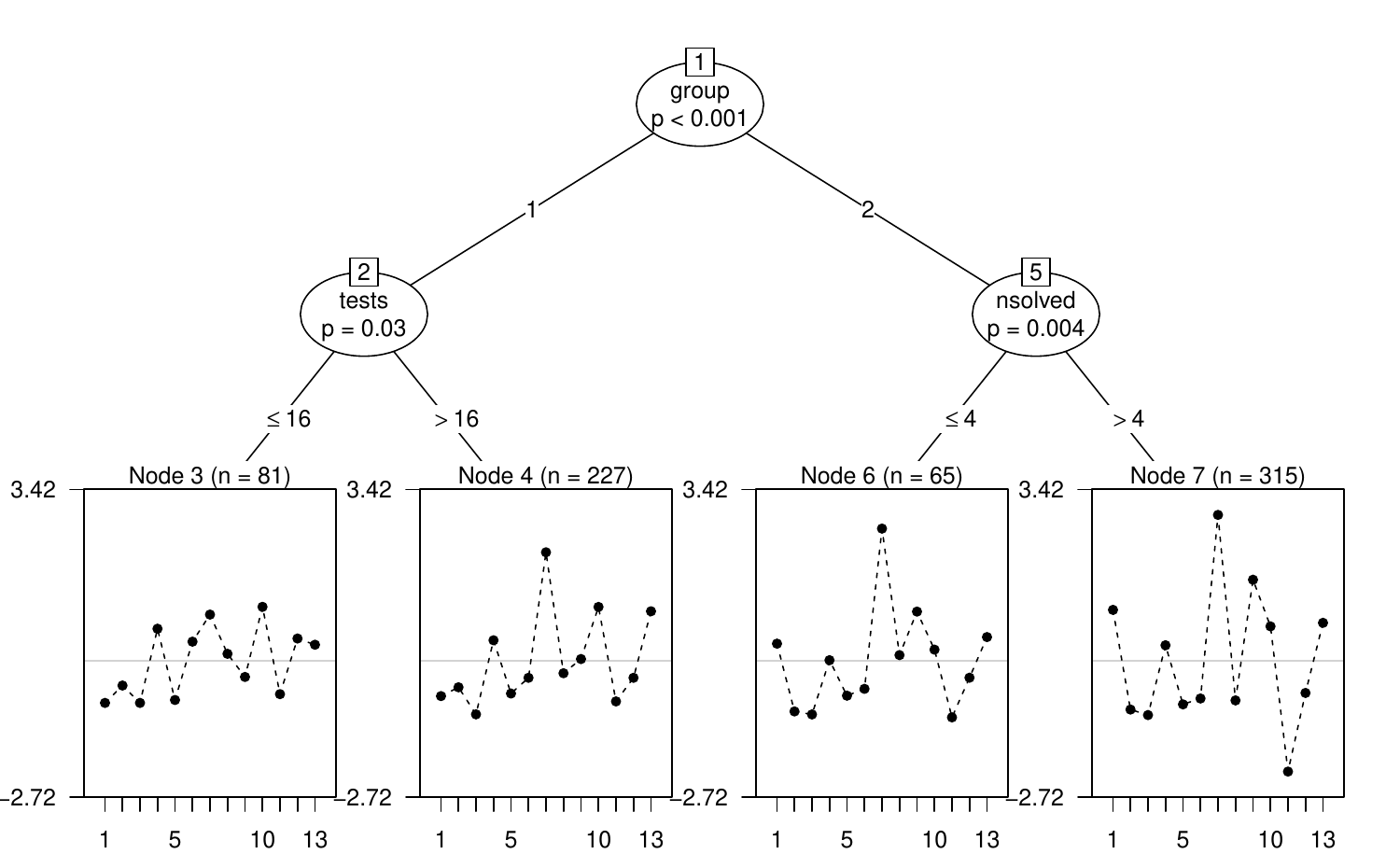}
\caption{Rasch tree for the mathematics exam data, detecting DIF with respect
to the exam group (1 vs.\ 2) and with respect to the mathematical abilities,
captured by the number of points in the online tests and the number of items solved
in the exam, respectively.}
\label{fig:raschtree}
\end{figure}

The resulting tree can be visualized with \code{plot(mrt)} and is depicted in
Figure~\ref{fig:raschtree}. The tree detects all of the subgroups we had considered
above in a data-driven way. First, the highest amount of DIF is found with respect
to the first vs.\ second group. The differences can be seen between the two item
profile plots on the left (nodes~3 and~4) and those on the right (nodes~6 and~7).
For example, in the second group the first item was much more difficult than the
subsequent items while it was somewhat easier in the first group. Similarly,
item~9 is easier than item~10 in the first group but more difficult in the second group.
Within both groups further evidence for DIF is found, splitting both groups
into a smaller subgroup of persons with lower mathematical abilities (nodes~3 and~6, respectively)
and a larger subgroup of persons with higher mathematical abilities (nodes~4 and~7, respectively).
In the first group this split is characterized by the number of points obtained
in the online tests during the semester while in the second group it is with respect
to the number of items solved correctly within the exam. Qualitatively, the change
in the item profiles is similar within both groups. For the students with higher
mathematical abilities there is a clear profile that distinguishes between
easier and harder items. In contrast, for the students with lower mathematics skills,
the item profile is very much ``dampened'' (especially in group~1), reflecting that for them all items are
almost equally easy or hard.

\subsection{Rasch mixture models}

Finally, we consider another possibility to assess measurement invariance in
Rasch models when there are no covariates available at all. This can be accomplished
using finite mixtures of Rasch models \citep{Rost:1990}. The idea is that there
are two or more subgroups or clusters with different item parameters but the
cluster membership is \emph{not} known. But it can be inferred using the EM
(expectation-maximization) algorithm which iterates between estimating the
parameters in all clusters (M-step) and then assigning each observation to
those cluster(s) it is most similar to (E-step).

In \proglang{R}, the package \pkg{flexmix} \citep{Leisch:2004,Gruen+Leisch:2008}
provides an modular implementation of finite mixture models which
\cite{Frick+Strobl+Leisch:2012} have interfaced in \pkg{psychomix} for IRT modeling with the
\pkg{psychotools} package. \cite{Frick+Strobl+Zeileis:2015} further extended
the functionality by considering enhancements for selecting the necessary
number of clusters and for appropriately modeling the cluster-specific
distribution of sum scores.

\begin{figure}[t!]
\includegraphics[trim=0 5 0 50, clip]{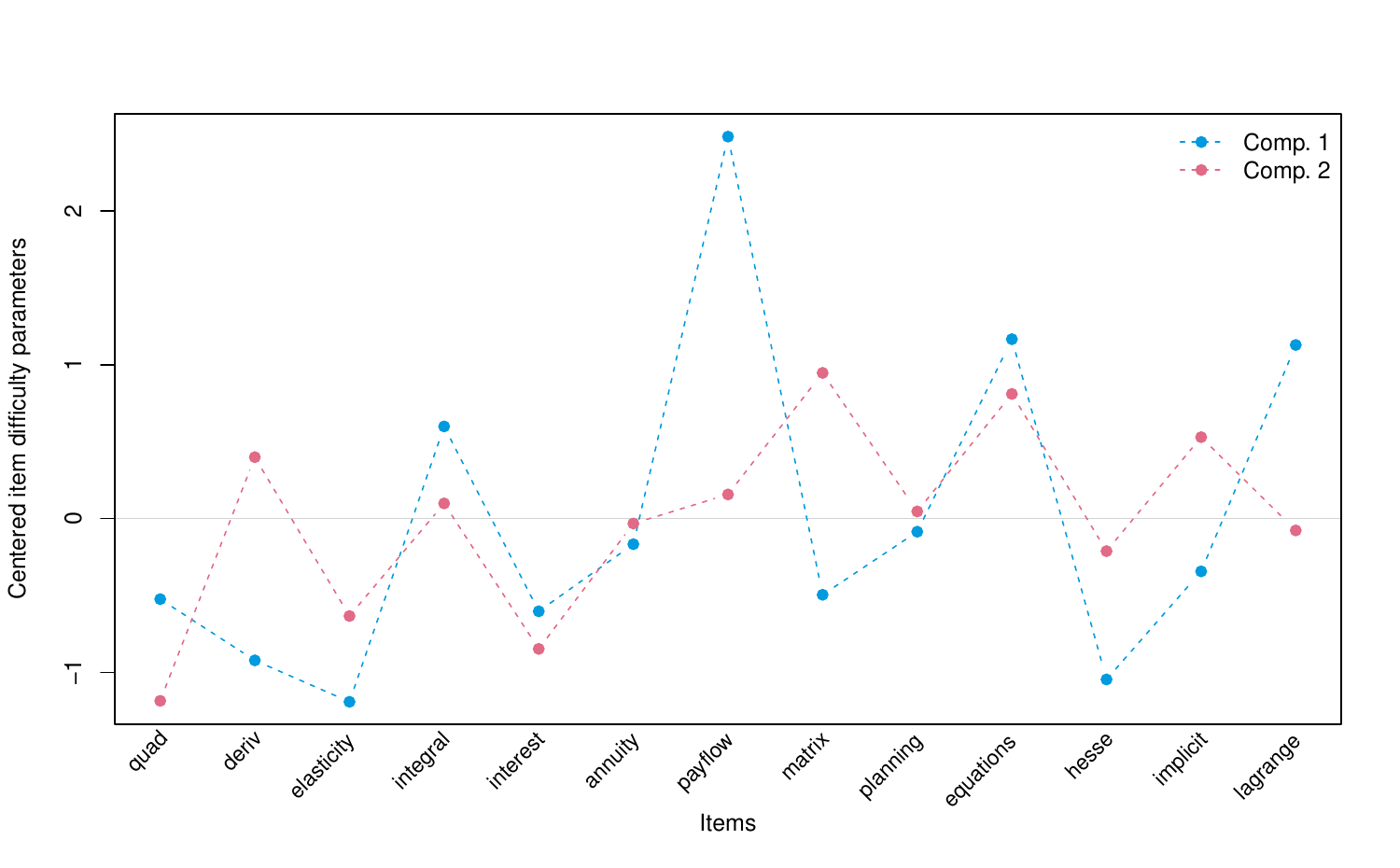}
\caption{Item profile plots for 2-component Rasch mixture model for the
students in group~1.}
\label{fig:raschmix}
\end{figure}

Here, we only illustrate the method briefly by asking whether it would be
possible to detect DIF in \code{group}~1 without knowing the relevant covariate
\code{tests}. We do so by directly fitting a 2-component mixture (without selecting
the number of clusters) with component-specific sum score distribution in
mean-variance specification.
\begin{Schunk}
\begin{Sinput}
R> library("psychomix")
R> mrm <- raschmix(mex1$solved, k = 2, scores = "meanvar")
R> plot(mrm)
R> print(mrm)
\end{Sinput}
\end{Schunk}
\begin{Schunk}
\begin{Soutput}
Call:
raschmix(formula = mex1$solved, k = 2, scores = "meanvar")

Cluster sizes:
  1   2 
235  73 

convergence after 73 iterations
\end{Soutput}
\end{Schunk}
The result of the Rasch mixture model is described by the \code{print()}
output above and the \code{plot()} output in Figure~\ref{fig:raschmix}.
Qualitatively, this provides the same insights as nodes~3 and~4 in
Figure~\ref{fig:raschtree}. There is a larger group that clearly discriminates
between the items that are relatively easy and those that are relatively
difficult. But there is also a smaller group whose item profile is very
much ``dampened'' and for whom all items are almost equally easy or
equally difficult.

The main difference is that here the Rasch tree is better at picking
up the subgroups because it can exploit the available covariates. Moreover,
the tree yields ``hard'' splits where a person is only in exactly one
subgroup of the tree. In contrast, in the finite mixture model, there
are only ``soft'' classifications so that each person is part of all
clusters but with different weights. See \cite{Frick+Strobl+Zeileis:2014}
for a comparisons of regression trees and finite mixture and what
their relative advantages and disadvantages are.

\section{Discussion} \label{sec:discussion}

This paper presents a hands-on tutorial for examining the results from
multiple-choice exams using IRT models along with methods for assessing
measurement invariance, in particular differential item functioning (DIF).
The flexible toolbox encompasses methods for detecting violations along
one covariate (tests), many covariates (trees), or no covariates (mixtures).
The covariates do not need to be dichotomized in advanced but can exploit
their different scales, be it continuous, ordinal, or categorical.

For the data set presented, results from an introductory mathematics exam, along with
a number of different covariates, the Rasch tree gives probably the
quickest overview of the underlying DIF patterns without having to specify
specific subgroups in advance.

The insights from this analysis had a number of policy implications for
the introductory mathematics exams at Universit\"at Innsbruck. First, exam
groups are avoided, if at all possible. Second, seemingly equivalent items
can function very differently if students focus their learning on well-known
parts of the item pool. This was the case for the first item (on optimization
of a quadratic function). Both question texts were very similar in terms
of the mathematical problem they described. However, one story was used
in the online tests during the same semester (group~2) and the other story was used
in the end-term exam from the previous semester (group~1) that all students had
access to. And it turned out that the students perceived the latter exercise
as much easier, possibly because they practiced with last semester's exam
more than with the online test exercises.

The analysis ties together a number of \proglang{R} packages, all of which
are freely available under the General Public License from the Comprehensive
\proglang{R} Archive Network (CRAN) at \url{https://CRAN.R-project.org/}.
\pkg{strucchange} provides an object-oriented implementation of the
score-based parameter instability tests. Model-based recursive partitioning
is available in \pkg{partykit} and model-based clustering with finite mixture
models in \pkg{flexmix}. Psychometric models that cooperate with
\pkg{strucchange}, \pkg{partykit}, and \pkg{flexmix}, are implemented in
\pkg{psychotools}, including various IRT models (Rasch, partial credit,
rating scale, parametric logistic), Bradley-Terry, and multinomial processing
tree models. The corresponding psychometric trees are in \pkg{psychotree}
and the psychometric mixture models in \pkg{psychomix}.

Finally, the \proglang{R} package \pkg{exams} provides flexible infrastructure
for conducting large-scale exams using randomized dynamic item pools either
in classical written exams (optionally with automatic evaluation) or in
online learning management systems (such as Moodle, OpenOlat, Canvas, Blackboard,
etc.). It is based on exercise templates with single- or multiple-choice items,
numeric exercises, closed and open-ended text questions, as well as combinations
of all of these. 

\section{Epilogue} \label{sec:epilogue}

This paper is dedicated to the memory of our friend and colleague Friedrich
``Fritz'' Leisch who died after serious illness in April 2024. Fritz contributed
to the work presented here in a number of ways: He co-developed the \pkg{strucchange}
package and some of the parameter instability tests which were originally geared
towards testing for structural changes in time series regressions. And it was only
somewhat later that we realized that we could adapt the same tests for recursive
partitioning in regression trees \citep{Zeileis+Hothorn+Hornik:2008}. While Fritz
was not a co-author of the latter work, he had nevertheless been an influence for it, mainly due to
his work on model-based clustering with finite mixture models \citep{Leisch:2004}.
The way that Fritz had set up a modular framework for plugging different
kinds of models into \pkg{flexmix} inspired us to establish a similar model-based
framework for recursive partitioning along with an object-oriented implementation.
Later on we came full circle by developing together the Rasch mixture models in
\pkg{psychomix} \citep{Frick+Strobl+Leisch:2012}. Finally, Fritz was also a
co-author of the \proglang{R}/\pkg{exams} infrastructure. He was the first who
had successfully extended our original implementation (which was geared towards
PDF output only) to XML exports for the Moodle learning management system. So
we joined forces to establish a flexible toolbox that can generate all kinds
of different exports from the same dynamic item pool.

Fritz will be dearly missed as colleague, collaborator, and friend.

\bibliography{fritz}

\end{document}